\documentclass[a4paper,conference]{IEEEtran}

\ifCLASSINFOpdf
  \usepackage[pdftex]{graphicx}
  \graphicspath{{../pdf/}{../figures/}}
  \DeclareGraphicsExtensions{.pdf,.jpeg,.png}
\else
 
\fi

\usepackage{bm}
\usepackage{xcolor, soul}
\usepackage{amsmath}
\usepackage{amsfonts} 

\ifCLASSOPTIONcompsoc
  \usepackage[caption=false,font=normalsize,labelfont=sf,textfont=sf]{subfig}
\else
  \usepackage[caption=false,font=footnotesize]{subfig}
\fi

\usepackage{lipsum}% http://ctan.org/pkg/lipsum
\usepackage{graphicx}% http://ctan.org/pkg/graphicx

\begin{document}

\title{Investigating Brain Connectivity with Graph Neural Networks and GNNExplainer}

\author{\IEEEauthorblockN{Maksim Zhdanov}
\IEEEauthorblockA{TU Dresden, HZDR \\
Dresden, Germany\\
Email: m.zhdanov@hzdr.de}

\and
\IEEEauthorblockN{Saskia Steinmann}
\IEEEauthorblockA{UKE \\
Hamburg, Germany\\
Email:  s.steinmann@uke.de}
\and
\IEEEauthorblockN{Nico Hoffmann}
\IEEEauthorblockA{HZDR\\
Dresden, Germany\\
Email:  n.hoffmann@hzdr.de}

}

\maketitle

\begin{abstract}

Functional connectivity plays an essential role in modern neuroscience. The modality sheds light on the brain's functional and structural aspects, including mechanisms behind multiple pathologies. One such pathology is schizophrenia which is often followed by auditory verbal hallucinations. The latter is commonly studied by observing functional connectivity during speech processing. In this work, we have made a step toward an in-depth examination of functional connectivity during a dichotic listening task via deep learning for three groups of people: schizophrenia patients with and without auditory verbal hallucinations and healthy controls. We propose a graph neural network-based framework within which we represent EEG data as signals in the graph domain. The framework allows one to 1) predict a brain mental disorder based on EEG recording, 2) differentiate the listening state from the resting state for each group and 3) recognize characteristic task-depending connectivity. Experimental results show that the proposed model can differentiate between the above groups with state-of-the-art performance. Besides, it provides a researcher with meaningful information regarding each group's functional connectivity, which we validated on the current domain knowledge. 
\end{abstract}

\IEEEpeerreviewmaketitle

\section{Introduction}
%What problem are we solving? Why is it important? How do we do it (EEG)? 
The human brain is a complex highly-connected neuronal network consisting of multiple regions. Those regions demonstrate neural synchronization when conducting a cognitive function. Throughout the past decade, the phenomenon has been studied by functional neuroimaging \cite{ONeill2018DynamicsOL}. It is argued that the co-activation of spatially distributed areas in the cerebral cortex can carry information about the way the brain works. In the case of brain pathologies, those networks are often altered due to the damaged brain structure \cite{Du2018ClassificationAP}. Hence, the analysis of such networks is of primary clinical interest as the source of biomarkers for a specific disease.

One particular example linked to abnormal co-activation of brain regions is schizophrenia \cite{Li2019DysconnectivityOM}. The field of neuroimaging has contributed drastically to understanding the illness by discovering alterations in the functional network associated with it. One frequent symptom of schizophrenia is auditory verbal hallucinations (AVH) - hearing voices without any actual external speaker. Even though schizophrenia has been studied extensively in the past decades, there is no consensus on the neural mechanisms underlying AVH \cite{Steinmann2017AuditoryVH}. According to recent findings \cite{PMID:17884165}, \cite{jardri2011}, the substantial similarity between the hallucinations and real voices is connected to alterations in a brain network related to speech and language processing. Steinmann et al. \cite{Steinmann2017AuditoryVH} demonstrated that the occurrence of AVH is related to altered synchronization of gamma-band connectivity while performing a dichotic listening task. 

\begin{figure*}[]
  \includegraphics[width=\textwidth]{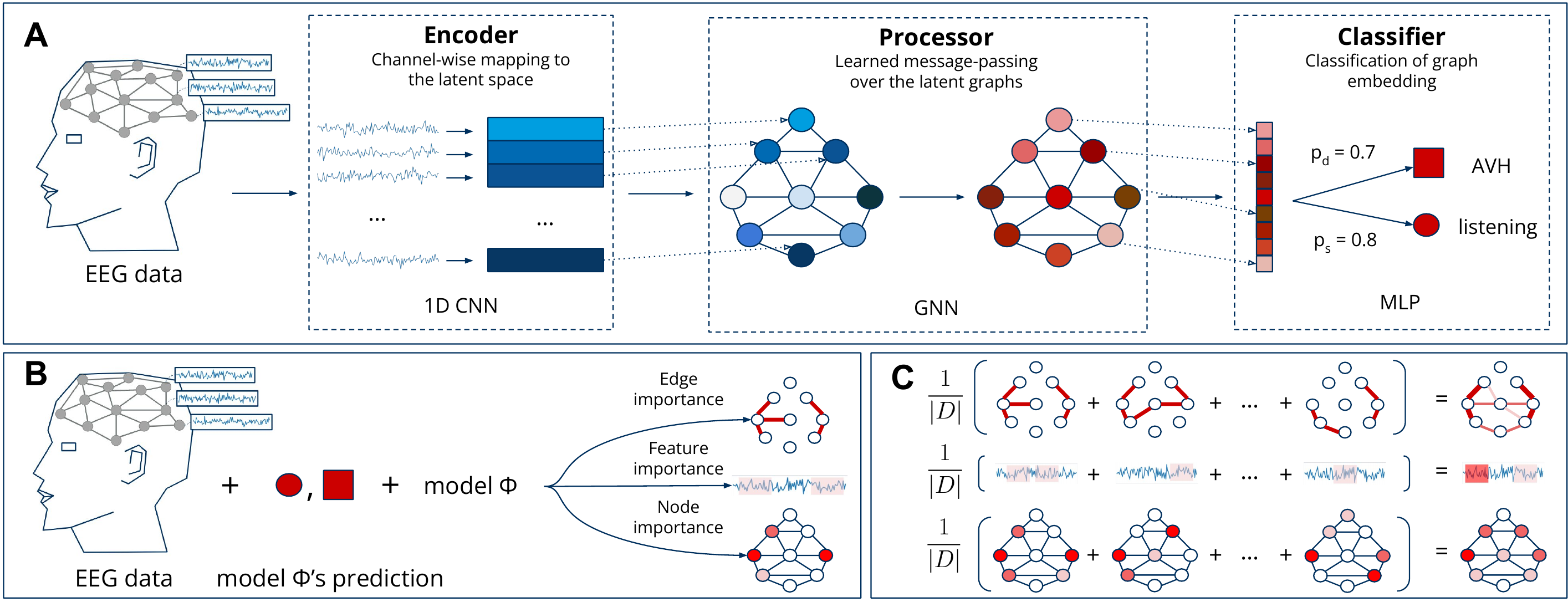}
  \caption{Scheme of the proposed approach. (A) Latent temporal features are computed with 1D CNN in a channel-wise manner. Afterwards, the latent feature vectors are mapped on a graph determined by the EEG setup. Multiple steps of message passing are further applied via graph neural networks to calculate the hidden representation of the graph. Node-wise max pooling is then applied to produce a graph embedding. The embedding is passed through an MLP to predict the graph's disorder and state classes. (B) GNNExplainer takes EEG data, a model and a prediction of the model to infer a subset of nodes (red nodes), a subset of edges (red edges) and a subspace of node features (transparent red area) that the model relied on while making the decision. (C) Class-wise explanations are computed for each edge as the frequency of being included in the explanation subgraph.}
  \label{fig:pipeline}
\end{figure*}

Electrophysiological methods such as electroencephalography (EEG) are widely employed in functional neuroimaging \cite{ONeill2018DynamicsOL}. They measure high-frequency oscillations with high temporal resolution, sufficient spatial resolution, and low cost. The general experimental setup for EEG consists of multiple interconnected electrodes located around a skull recording brain electrical activity. Each electrode covers a region of the cerebral cortex yielding a mapping between the electrode space and the cortex space. Thus, one can recover the region co-activation from EEG data with the mapping.

Since both connectivity networks and electrode setups can be represented as graphs, we utilize the framework of graph neural networks (GNN) to operate in the domain effectively. GNN perform convolution on graphs and compute graph embedding via learned message passing. GNN is applied to many areas of research, including neuroscience \cite{Arslan2018GraphSM}, physics simulation \cite{SanchezGonzalez2020LearningTS} and drug design \cite{Gaudelet2021UtilizingGM}. The broad applicability of GNN led to the development of approaches allowing one to explain predictions made by GNN-based models. Those approaches include attention-based \cite{Velickovic2018GraphAN}, adaptation of saliency maps \cite{Zeiler2014VisualizingAU} and mutual information-based such as GNNExplainer \cite{ying2019gnnexplainer}. 

In this paper, we introduce graph neural networks to EEG data analysis. Our goal is to discover brain regions that are activated when performing a dichotic listening task for three groups of people: healthy controls (HC group), schizophrenia patients (SZ group) and people who have schizophrenia with AVH (AVH group). To do so, we represent EEG data consisting of EEG electrodes and EEG recordings with graphs. A GNN-based classifier is then used to differentiate between the groups above. We further perform inference tasks with GNNexplainer. To the best of our knowledge, this is the first work demonstrating the applicability of GNNExplainer to EEG data. The main contributions of this work are as follows:
\begin{enumerate}
    \item We developed a GNN-based classifier that outperforms baseline approaches in the EEG-data classification task.
    \item We provide a scheme to perform inference and validate the model with GNNExplainer.
    \item We validated the resulting explanations on evidence from neurobiological studies showing the relevance of the proposed method.
\end{enumerate}

\section{Related Works}
%review some state-of-the-art (SOTA) ML methods for analyzing EEG data. Wrap up section with a couple of sentences about our approach differs from sota.
Our research has been conducted in two directions: a) developing a classifier receiving EEG data as input; b) identifying regions of interest based on predictions made by the classifier. In this section, we provide a brief review of each line. 
\subsection{Classification of EEG measurements with Deep Learning}
Deep learning gained significant attention in EEG data classification tasks in the past years. It was applied to seizure prediction \cite{Eberlein2018ConvolutionalNN}, sleep phase detection \cite{Yildirim2019ADL}, disorder detection (e.g. Alzheimer \cite{Bi2019EarlyAD}) and many other areas. As a rule, the classification pipeline for EEG data removes artefacts, extracts features, and then performs prediction. The most popular choice for a classifier's architecture is convolutional neural networks (CNN), followed by deep belief networks and recurrent neural networks. CNN-based models typically achieve the best performance \cite{Craik2019DeepLF}. The advantage of convolution-based networks is arguably due to their ability to extract temporal patterns from EEG signals \cite{Bronstein2021GeometricDL}. We leverage the advantages of CNNs, including them in our classifier to compute temporal features from EEG data to further pass through graph neural networks.
\subsection{Identification of task-related regions of interest}
There has been substantial interest in using GNN in functional neuroimaging recently. However, most research was done by applying GNN to extract patterns from functional magnetic resonance imaging (fMRI) data. Pioneering work has been done by Arslan et al. \cite{Arslan2018GraphSM} where graph saliency maps were calculated to explain how a GNN-based classifier tells apart males from females. It was followed by the work of \cite{Kim2020UnderstandingGI} who introduced importance score based saliency maps to find biomarkers associated with an autism spectrum disorder. A different approach was applied by Li et al. \cite{Li2020BrainGNNIB} who developed a pooling layer highlighting salient regions of interest. While fMRI is a classic method to investigate functional connectivity, it does not possess temporal resolution high enough to be applied to the analysis of network dynamics \cite{ONeill2018DynamicsOL}.
On the other hand, EEG is a low-cost, high-resolution method that directly measures neuronal oscillations. It is also significantly cheaper and more accessible compared to fMRI. For these reasons, we introduce a model that can work in the domain of EEG data and gather neurological biomarkers from it.

\section{Methods}
The proposed EEG analysis pipeline consists of 3 steps (see Fig. 1). The connectivity of electrodes in the experimental systems defines the graph. Each node of this graph corresponds to an electrode and has the EEG recording attached. We further calculate temporal local features for each electrode separately with 1D convolutional neural networks (see \ref{section:model}, Encoder). Afterwards, a graph neural network with multiple message-passing rounds calculates local spatial features for each electrode (see section \ref{section:model}, Processor). Next, we calculate the embedding of the resulting latent graph via global max pooling. The embedding is fed to the classifier that returns a probability of the input graph to belong to a disorder class and a state class (see section \ref{section:model}, Classifier). We explicitly separate disorder labels and state labels as we assume them to be independent. 

Explanations of the model predictions are further calculated via GNNExplainer for each subject and aggregated to get a class-level explanation. As a result, an average importance score is calculated for each electrode, pair of electrodes and time point. Consequently, we employ known mapping between electrodes and brain regions to infer the functional connectivity for each disorder group.

\subsection{Notation} 

The experimental setup of EEG recording can be represented as a graph with the set of nodes $V$ being electrodes and the set of edges $E$ being connections between the electrodes. Let $X = (x_1,x_2,...,x_{N})^T \in \mathbb{R}^{N \times d}$ be a set of EEG recordings obtained from $N$ channels. We thus define X as the set of node features where $x_v \in \mathbb{R}^d$ for all $v \in V$, $d$ is the length of a recording. Each set of EEG recordings is associated with 2 labels: $state \: s \in S = \{resting, listening\}$ and $diagnosis$ $d \in D = \{HC, SZ, AVH\}$ where $HC$ denotes healthy control group, $SZ$ - schizophrenia patients, $AVH$ - group suffering from schizophrenia followed by AVH. Hence, for each experiment we receive a graph $G = (V, E, X) \in \mathcal{G}$, a corresponding diagnosis $d \in D$ and a state $s \in S$. We will further refer to a graph $G$ as a tuple $(A, X)$ where $A$ is a binary adjacency matrix derived from $V$ and $E$.

\subsection{Classification model} 
\label{section:model}
The model $\Phi \in \mathcal{F}$ learns conditional distribution $P(d,s|G)$ in 3 steps (see Fig. \ref{fig:pipeline}A):

\textbf{Encoder: $ \bm{\mathbb{R}^{d} \rightarrow \mathbb{R}^{d_{latent}}}$} maps node features from EEG space to a latent space of dimension $d_{latent}$. It does so in a node-wise manner yielding $X_{enc} = (\hat x_1, \hat x_2,..., \hat x_{N})^T \in \mathbb{R}^{N \times d_{latent}}$ where $\hat x_v = Encoder(x_v) \: \forall v \in V$. The encoder is implemented as a stack of 1D CNN extracting temporal features from an EEG signal. As output, a latent representation of an input graph $G_{enc} = (A, X_{enc}) \in \mathcal{G}_{enc}$ is returned. To exploit the symmetries of EEG recordings, we make the mapping to the latent space translation and permutation equivariant. The use of convolutional neural networks guarantees translation equivariance \cite{Bronstein2021GeometricDL}. Permutation equivariance is preserved by independently applying 1D CNN for each node, thus holding $Encoder(P \cdot X) = P \cdot Encoder(X)$ for any permutation matrix $P$. As a result, the nodes of a latent graph $G_{enc}$ carry information about the same temporal patterns occurring in EEG data, e.g., synchronised electrical response to a stimulus.

\textbf{Processor: $ \bm{\mathcal{G}_{enc} \rightarrow \mathcal{G}_{proc}}$} computes a hidden representation of a graph with preserved graph structure but learned spatio-temporal features. Those features are calculated via the interaction of neighbouring nodes in multiple rounds of learned message-passing \cite{morris2020weisfeiler}:
\begin{equation}
    \label{eq:gnn}
    \begin{cases}
      h^{0}_{v_i} = \hat x_{v_i} \\
      h^{n}_{v_i} = ReLU (\Theta_1 \cdot h^{(n-1)}_{v_i} +
        \Theta_2  \cdot \sum\limits_{v_j \in \mathcal{N}(i)} h^{(n-1)}_{v_j})
    \end{cases}
\end{equation}
where superscript denotes the number of message-passing rounds, $\mathcal{N}(i)$ is the first-order neighbourhood of the node $v_i$; $\Theta_1, \Theta_2$ are matrices optimized during training. The function returns a hidden graph $G_{proc} = (A, H) \in \mathcal{G}_{proc}$ as output where $H = (h^{N_{mp}}_1,h^{N_{mp}}_2,...,h^{N_{mp}}_{N})^T \in \mathbb{R}^{N \times d_{hidden}}$, $N_{mp}$ - total number of message passing rounds. We construct the mapping to preserve permutation equivariance to account for the symmetry of EEG data defined on the graph domain \cite{Bronstein2021GeometricDL}. Thus, we calculate local spatial features with locality described by the number of message-passing steps.

\textbf{Classifier: $ \bm{\mathcal{G}_{proc} \rightarrow D \times S}$} calculates the probabilities of a hidden graph to belong to a particular diagnosis class $d$ and a state class $s$. First, it applies global max pooling to compute the embedding of the graph $\Vec{h}$ from its spatio-temporal node feature matrix $H$. Next, an MLP takes the embedding and returns the probability of belonging to a particular state class and a particular disorder class:

\begin{equation}
    \label{eq:mlp}
    \begin{cases}
      p(s|H) = softmax(W_s \cdot ReLU ( W \cdot \Vec{h} +  b) + b_s) \\
      p(d|H) = softmax(W_d \cdot ReLU ( W \cdot \Vec{h} +  b) + b_d)
    \end{cases}
\end{equation}
where $\Vec{h}$ is the embedding of the hidden graph $G_{proc}$, $W, b, W_d, W_s, b_d, b_s$ are matrices optimized during training.

\subsection{Explanation pipeline}
After the model $\Phi$ is trained, we are interested in receiving explanations for its predictions. Specifically, for any prediction of $\Phi$, we want to obtain a subgraph of an input graph $G = (A, X)$ that was the most influential.

\textbf{GNNExplainer: $ \bm{(\mathcal{F},\mathcal{G},D,S) \rightarrow \mathcal{G}_{sub} \subseteq \mathcal{G}}$ } searches for a subgraph $G_{sub} = (A_{sub}, X_{sub})$ of an input graph $G$ such that mutual information between the model $\Phi$'s prediction and the distribution of possible subgraphs is maximized \cite{ying2019gnnexplainer}: 
\begin{equation}
    \operatorname*{argmax}_{G_{sub} } MI(d, s, G_{sub} ) = H(d,s) - H(d, s|G=G_{sub})
\end{equation}

Computationally efficient version of the objective is formulated as described in \cite{ying2019gnnexplainer}. One learns masks for an adjacency matrix $A_i$ and a feature input matrix $X_i$ of a graph $G_i$ via gradient descent. Precisely, for each graph $G_i = (A_i, X_i)$ the following objective is maximized:
\begin{equation}  \label{eq:GNNExplainer_obj}
    \operatorname*{max}_{M_1, M_2} \: log \: p_{\Phi} (D = d_i, S = s_i | A = A_i \odot M_1, X = X_i \odot M_2)
\end{equation}
where $\odot$ denotes element-wise multiplication, $M_1, M_2$ - learnable masks for an adjacency matrix and a feature input matrix, $\Phi$ is a trained model.

We further reformulate the objective (\ref{eq:GNNExplainer_obj}) to adapt it for the task of functional connectivity examination. Namely, we are interested in evaluating the co-activation of electrodes, information pathways and time of the response to stimuli for different groups of people. We solve three tasks tackled separately from one another (see Fig. \ref{fig:pipeline}B).

\subsubsection{Node co-activation}

Our task is to define the nodes which a learned model $\Phi$ relies on when making a prediction for a single graph $G_i$. To do so, we adapt the objective (\ref{eq:GNNExplainer_obj}) and learn a mask for a feature input matrix $X_i$:
\begin{equation}
    \operatorname*{max}_{M_V} \: log \: p_{\Phi} (D = d_i, S = s_i | X = X_i \odot \sigma(M_V))
\end{equation}
where $\odot$ denotes element-wise multiplication, $M_V \in \mathbb{R}^{|V|}$ is a mask that masks out nodes of an input graph, $\sigma$ is the sigmoid function to map the mask to $[ 0, 1 ]^{|V|}$. It can be seen as masking out rows of an input feature matrix $X$ that do not significantly contribute to the final prediction of the model. Additionally, we minimize element-wise entropy for the mask $M_V$ to be either 0 or 1 and the number of non-zero elements of $M_V$ to penalize the large explanation size. The weights of nodes given by $\sigma (M)$ are hence binary and denote whether the model activates a node or not while making a prediction. Consequently, we assign those activation labels to the corresponding brain regions. We argue that the electrical activity of those regions is distinct when performing a particular function for a given recording.
\subsubsection{Information pathways} One can also search for connections in the edge set $E$ that the model utilizes while performing learnable message-passing. We learn a mask of a single graph adjacency matrix $A_i$ by adapting the objective (\ref{eq:GNNExplainer_obj}):
\begin{equation}
    \operatorname*{max}_{M_A} \: log \: p_{\Phi} (D = d_i, S = s_i | A = A_i \odot \sigma(M_A))
\end{equation}
where $\odot$ denotes element-wise multiplication,  $M_A \in \mathbb{R}^{|E|}$ is a mask of an adjacency matrix, $\sigma$ is the sigmoid function to map the mask to $[ 0, 1 ]^{|E|}$. As in the previous task, element-wise entropy and the number of non-zero elements of the mask are minimized. Hence, the framework seeks for the subset of edges $E_{sub}$ that is used by the model $\Phi$'s Processor during message-passing steps. As a result, the resulting subset of edges would correspond to brain connections through which the important information "flows".

\subsubsection{Time of the response to stimuli} We can also modify the objective such that the framework learns the subspace of input node feature space that is crucial for the model's predictions. It is equivalent to learning a mask for a feature input matrix $X_i$ by adapting the objective (\ref{eq:GNNExplainer_obj}):
\begin{equation}
    \operatorname*{max}_{M_T} \: log \: p_{\Phi} (D = d_i, S = s_i | X = X_i \odot \sigma(M_T))
\end{equation}
where $\odot$ denotes element-wise multiplication, $M_T \in \mathbb{R}^{d}$ is a mask that masks out input feature dimensions, $\sigma$ is the sigmoid function to map the mask to $[ 0, 1 ]^{d}$. We also minimize element-wise entropy and the number of non-zero elements of the mask. Since input node feature space is a space of EEG signals, the resulting subspace would identify regions of a signal crucial for the model. Hence, the framework highlights the most informative time spans to differentiate functional brain activity for each disorder.

\textbf{Class-level explanations.}
We now know how to calculate explanation masks for each instance of a state class $s$ and a disorder class $d$ in the dataset (see Fig. \ref{fig:pipeline}C). We further aggregate those masks to obtain class-level explanations. We do so by computing the average of the learned masks $\bar{M}_{d,s} = \frac{1}{|D_{d,s}|} \sum_i M^i_{d,s}$, where $M \in \{M_V, M_T, M_A\}$. Hence, for a given combination of a disorder $d$ and a state $s$, we quantify 1) the activation frequency for each node with $\bar{M}_V$, 2) the activation frequency of a time point with $\bar{M}_T$ and 3) the frequency of an edge of being included in the explanation network with $\bar{M}_A$.

\section{Experimental Details}
\subsection{Experimental data}
The study included 29 patients diagnosed with schizophrenia and 52 healthy controls. Every participant was right-handed. Among 29 schizophrenia patients, 14 subjects belonged to the AVH group. Each participant was given one of six different syllables (/ba/, /da/, /ka/, /ga/, /pa/, /ta/) for 500 ms simultaneously to each ear after 200 ms silence period. The EEG recording was conducted with 64 electrodes, including 4 EOG channels, to monitor eye movements during the task. For each subject, the experiment was conducted multiple times (number of trials for AVH: $68.23 \pm 19.43$; SZ: $68.76 \pm 14.79$ and HC: $71.19 \pm 12.93$). The data was filtered from 20 to 120 Hz at the preprocessing step. We further downsampled signals to 500 Hz according to a protocol described in \cite{Steinmann2017AuditoryVH}. Afterwards, we took the common average with further re-referencing of all channels. Besides, muscle and visual artefacts were identified and removed. Each recording was then divided into a resting state (first 200 ms with no syllable given) and a listening state (initial 200 ms when syllables were presented). One can find detailed information on data acquisition and preprocessing in \cite{Steinmann2017AuditoryVH}.

To create a dataset for training models, we use the inter-patient paradigm, i.e. data from the same subject does not appear simultaneously in training, validation and test datasets. Thus, we randomly select two subjects from each group to form the validation dataset and one subject to form the test dataset. Data of the remaining subjects are randomly sampled to create the training dataset with an equal number of data points for each disorder class and state class. As a result, training/validation/test datasets contain $10290/1512/852$ samples.

\subsection{Implementation details}
\textbf{Encoder implementation}. The encoder uses 1D CNN consisting of $N_{cnn}$ convolutional layers with a kernel size of 5. $N_{cnn}$ is a hyperparameter of the model. Each convolutional layer is followed by ReLU non-linearity. As a result, an EEG recording is mapped into a latent space with dimension $d_{latent}$ which is also a hyperparameter of the model.

\textbf{Processor implementation}. The processor consists of $N_{mp}$ graph convolution layers with identical structures (eq. \ref{eq:gnn}). The number of message-passing steps $N_{mp}$ and hidden space dimension $d_{hidden}$ are hyperparameters of the model.

\textbf{Classifier implementation}. Classifier comprises two parts: graph embedding calculation and classification itself. We apply global max pooling to a node feature matrix $H$ in the first step. The embedding is further fed to an MLP to predict the probability for the graph to belong to a disorder $d$ and state $s$ (eq. \ref{eq:mlp}).

\textbf{Baseline models}. We compare our model with two deep learning architectures that are widely used in the field: EEGNet \cite{Lawhern2021EEGNetAC}, and ShallowNet \cite{Schirrmeister2017DeepLW}. See implementation details in Appendix. We optimize models the same way we optimize the GNN model (see section \ref{opt}).

\subsection{Training \& Explaining details}
\label{opt}
\textbf{Optimization procedure.} 
Parameters $\theta$ of every model are optimized via minimizing binary cross-entropy (BCE) loss at every step of training, i.e. $\mathcal{L}(p(s,d|G), s, d; \theta) = BCE(p(s,d|G), (s,d))$, for each graph in a mini-batch. We represent label $d$ as a two-dimensional binary vector $d = \{ has\_schizophrenia,  has\_avh \}$ (thus, HC = [0,0], SZ = [1,0], AVH = [1,1]). We use Adam optimizer with a learning rate of $10^{-4}$. The training was performed in mini-batches of size 32. 

\textbf{Explanation procedure.} We randomly select $500$ instances of each combination $d,s$ of class variables for each explanation task. The corresponding objective is maximized separately for each task with a learning rate of $10^{-3}$ for $2 \cdot 10^4$ epochs.

\section{Results \& Discussion}
We optimized hyperparameters of GNN and baseline models to infer the architecture that performs the best on the test dataset for each model. We compared the accuracy of diagnosis classification, the accuracy of state classification, and overall accuracy, which is their average. (see Table \ref{table:hypers}). We found that the performance of our model is robust for the choice of hyperparameters. Our model consistently and significantly outperforms baseline models in the diagnosis ($p < 10^{-5}$) and overall classification ($p < 10^{-4}$). EEGNet and ShallowNet demonstrated slightly higher state classification accuracy than the GNN model, yet they could not perform diagnosis prediction as good. It can be explained by the fact that the GNN-based model calculates Spatio-temporal features respecting the permutation symmetry of EEG data defined on the graph. Conversely, it is not the case for both baseline models, which use spatial convolutions that violate the symmetry.

To obtain the most reliable explanations, we further use the architecture that achieved the highest accuracy score on the test dataset ($N_{cnn} = 3, N_{mp} = 1, d_{latent} = 32, d_{hidden} = 64$).  

%p-value on accuracy s = 0.09
%p-value on accuracy d = 0.0001 
%p-value on accuracy = 0.03 
%p-value on f1 = 0.04
%p-value on recall = 0.03
%p-value on precision = 0.052

\begin{table}[]
\caption{Comparison of GNN-based model (ours) to baseline models in terms of accuracy on the test dataset. For each column, we highlight a row whose value is significantly higher than that of others.}
\centering
\scalebox{0.96}{
\begin{tabular}{|c|c|c|c|}
\hline
Model & $accuracy$ $d$ & $accuracy$ $s$ & $\overline{accuracy}$ \\ \hline
GNN(\textbf{Ours}) & $\mathbf{0.627 \pm 0.025}$ & $0.593 \pm 0.018$ & $\mathbf{0.610 \pm 0.015}$ \\ \hline
EEGNet \cite{Lawhern2021EEGNetAC} & $0.426 \pm 0.104$ & $0.663 \pm 0.059$ & $0.545 \pm 0.050$ \\ \hline
ShallowNet \cite{Schirrmeister2017DeepLW} & $0.433 \pm 0.050$ & $0.638 \pm 0.026$ & $0.536 \pm 0.029$\\ \hline
\end{tabular}}
\label{table:hypers}
\end{table}

\subsection{Node co-activation} \label{sec:node_co_ac}
\label{node}
\begin{figure}
  \includegraphics[width=0.48\textwidth]{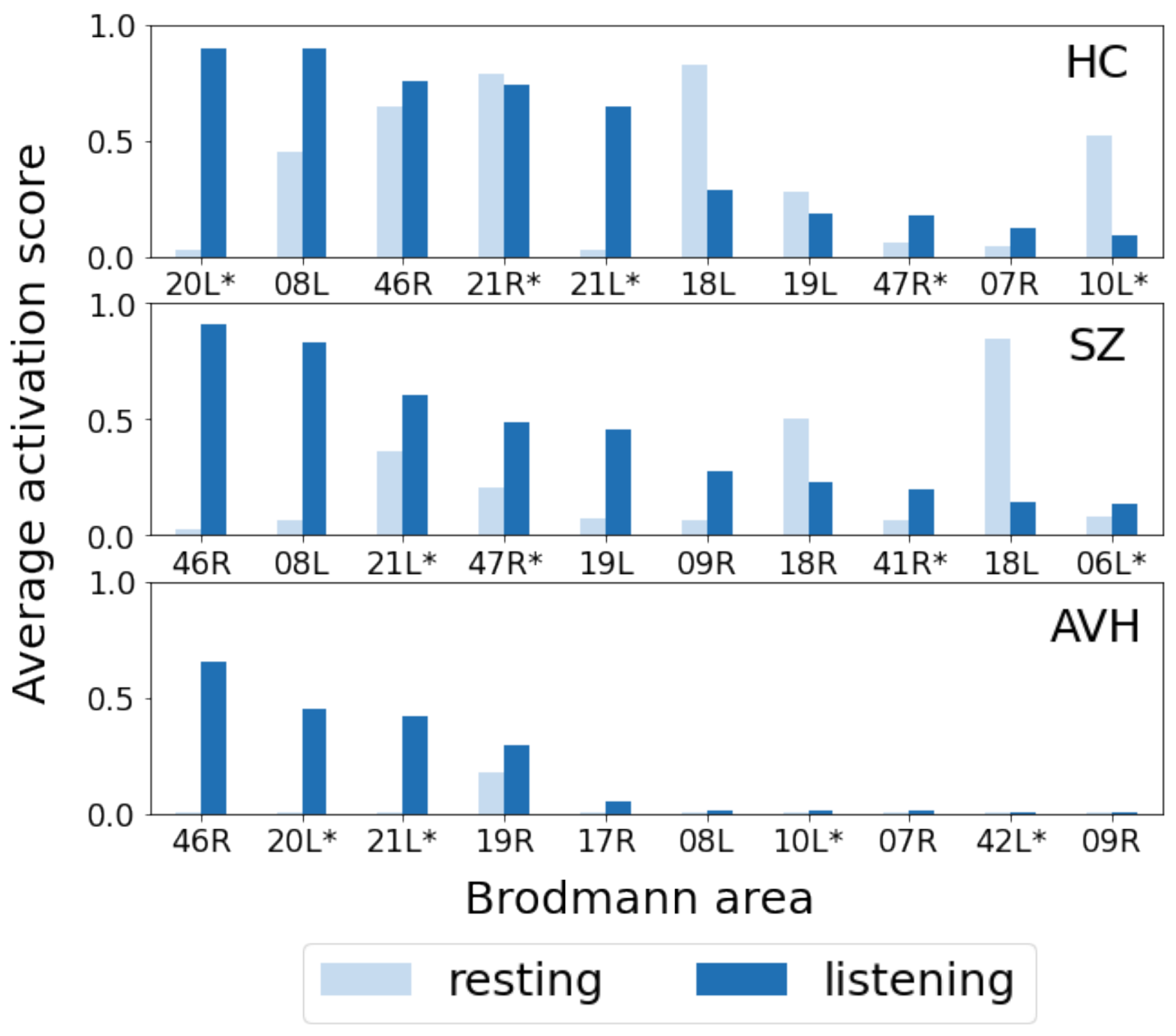}
  \caption{Brodmann area activation during listening and resting for every disorder group: healthy people (Top), schizophrenia patients (Middle), schizophrenia patients with AVH (Bottom). For each group, 10 Brodmann areas are taken that have the highest average activation scores. Bars show an average activation score. Brain regions that are found to be involved in auditory tasks for healthy people are denoted with $^{\ast}$.}
  \label{fig:node_scalar}
\end{figure}

\begin{figure*}[]
  \includegraphics[width=0.98\textwidth]{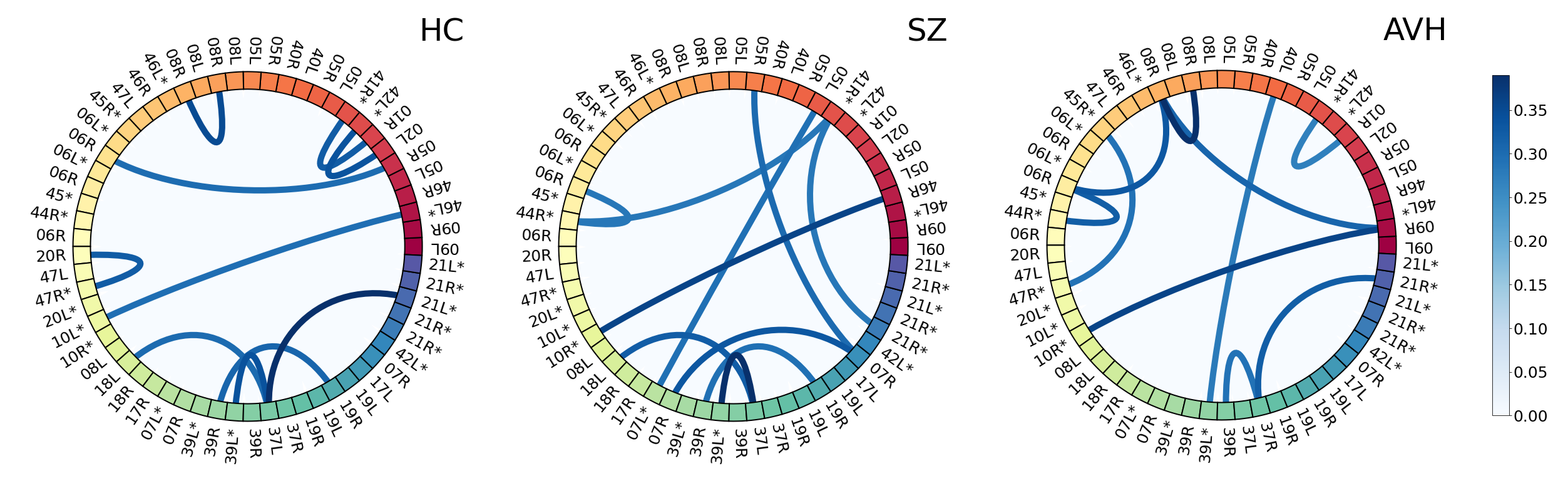}
  \caption{Connection's frequency of being activated during message-passing for every disorder group: healthy people (Left), schizophrenia patients (Middle), schizophrenia patients with AVH (Right). For each group, ten connections are taken that have the highest listening importance scores. Brodmann areas that are found to be involved in auditory tasks for healthy people are denoted with $^{\ast}$.}
  \label{fig:edge_scalar}
\end{figure*}

Disorder-specific node activation scores can be found in Fig. \ref{fig:node_scalar}. Comparing listening maps to resting maps for the healthy control group, one can see that the highest difference in activation scores is achieved for Brodmann areas located in the temporal cortex: BA$20L$, BA$21L$. Those areas encompass the lateral temporal region of the brain, which is involved in language processing tasks \cite{ba20, ba21}. Overall, the model's regions to distinguish between listening and resting for the HC group belong to the left hemisphere: BA$20L$, BA$21L$, BA$08L$, BA$18L$. It is not surprising that the map highlights the left hemisphere since the auditory speech-related function is left-lateralized for right-handed people. 

Activation maps for SZ and AVH groups feature BA$46R$ as the region with the highest difference between listening and resting activation scores. For healthy people, the left homologue of the area, BA$46L$, is known to be incorporated in phoneme processing \cite{ba46l}. It can be seen as an alteration since the model does not utilize the region to differentiate states for healthy people (equally high activation score for both resting and listening). Overall, there is no left hemisphere superiority for both groups compared to the healthy group, i.e. the model highlights Brodmann areas in both hemispheres. 
%This fact might support the hypothesis about reduced right-ear advantage \cite{Steinmann2017AuditoryVH} which manifests itself in the disappearance of left hemisphere superiority during dichotic listening for people with schizophrenia. 

\subsection{Information pathways}

Fig. \ref{fig:edge_scalar} shows which connections in the computation graph the model relies on the most when predicting a state or a disorder. One can see that the model utilizes edges that connect brain regions involved in speech processing for healthy people \cite{ba20}. In most cases, the information goes either to or from an auditory processing region. It indicates that the model gathers spatial information meaningfully at the processor step. The edge activation maps for the SZ and AVH groups are similar to those for healthy people. However, most of the edges depicted in Fig. \ref{fig:edge_scalar} (9 out of 10) for these groups correspond to connections between brain regions located in the right hemisphere. This is consistent with node co-activation patterns indicated by the model (see section \ref{sec:node_co_ac}). It may indicate the absence of left hemisphere superiority which was recently connected to the emergence of AVH \cite{Steinmann2017AuditoryVH}. 

\subsection{Time of the response to stimuli}
\begin{figure}[htpb]
  \includegraphics[width=0.48\textwidth]{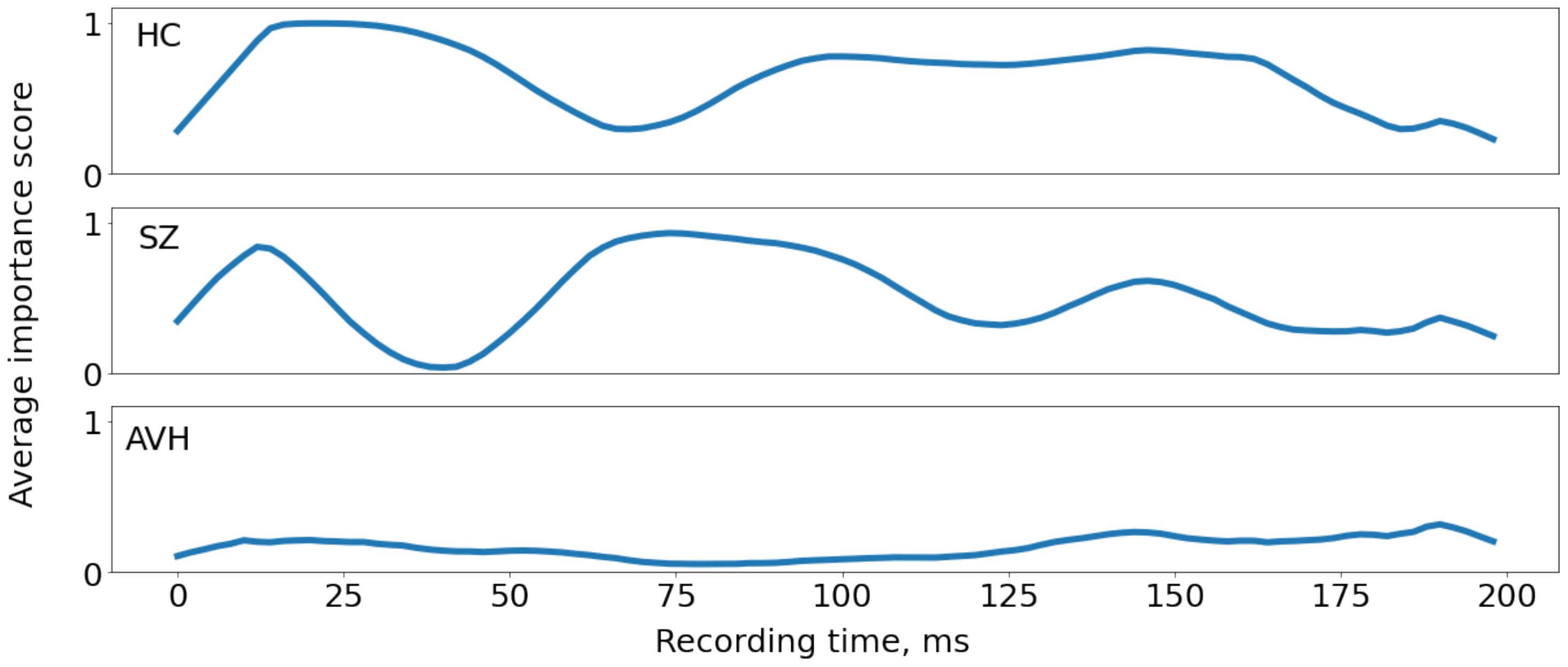}
  \caption{Average time point activation (blue) for listening for every group: healthy people (Top), schizophrenia patients (Middle), schizophrenia patients with AVH (Bottom).}
  \label{fig:node_vector}
\end{figure}

We have also adapted GNNExplainer's framework to highlight regions of EEG recordings that are informative for the model (see Fig. \ref{fig:node_vector}). The model consistently pays attention to the 20-50 ms and 90-170 ms periods during listening for healthy people. These periods correspond to Na/Pa, and N100 - auditory evoked potentials peaking between 20 to 30 ms and between  80 and 120 ms \cite{Ford2012NeurophysiologicalSO}. A slightly different pattern can be seen for the SZ group, which map indicates emphasizing No/Po and N100 evoked potentials. However, for the AVH group, the model does not consistently pay attention to a particular period. That could be an outcome of the reduced evoked potential, which is one of the most often replicated findings for patients who have schizophrenia \cite{Edgar2008SuperiorTG, Ford2012NeurophysiologicalSO}.

\section{Summary}
The paper introduced an end-to-end graph neural network to analyze EEG data. It reliably distinguishes schizophrenia patients based on EEG recording with and without auditory verbal hallucinations. The model respects symmetries of EEG recordings, namely the translational symmetry of EEG recordings and the permutational symmetry of EEG setup. Consequently, the proposed classifier outperforms baseline methods such as EEGNet and ShallowNet in classification accuracy. We further use the framework of GNNExplainer to identify salient electrodes, connections between electrodes and time regions of EEG recordings that the model relies on while making its predictions. To demonstrate the reliability of the explanations, we compared them with neuroscientific knowledge for the healthy people group. The model relies mainly on the brain regions incorporated in auditory processing for this group. Those regions and connections between healthy people were primarily located in the left hemisphere, which was not the case for schizophrenia patients. Finally, the model relies the most on the periods corresponding to auditory evoked potentials when predicting brain function. 

\newpage
\bibliographystyle{main}
\bibliography{main}

\end{document}

\appendices
\section{Implementation details}
We optimized hyperparameters of each model with respect to overall classification accuracy on the test dataset (see Table \ref{tab:architecture} for GNN model). 

\begin{table}[htpb]
    \centering
    \caption{Architecture of GNN model. Hyperparameters are optimized with respect to performance on the test dataset.}
    \label{tab:architecture}
    \begin{center}
\begin{tabular}{cc}
    \hline
    \textbf{Encoder} architecture \\ \hline
    Input $61 \times 100$ image \\ 
    Conv1D $32 \times 1 \times 31$ \& BatchNorm \& ReLU \\ 
    MaxPooling1D (kernel size $5$, stride $2$) \\
    Conv1D $32 \times 32 \times 31$ \& BatchNorm \& ReLU \\ 
    MaxPooling1D (kernel size $5$, stride $2$) \\
    Conv1D $32 \times 8 \times 31$ \& BatchNorm \& ReLU  \\ 
    Output $61 \times 32$ image \\
    \hline
    \end{tabular}
    \end{center}
\begin{center}
    \begin{tabular}{cc}
    \hline
    \textbf{Processor} architecture \\ \hline
    Input $61 \times 32$ image, $61 \times 61$ adjacency matrix \\ 
    GraphConv $(32,64)$ \& BatchNorm \& ReLU \\
    Output $61 \times 64$ image \\
    \hline
    \end{tabular}
\end{center}
\begin{center}
    \begin{tabular}{cc}
    \hline
    \textbf{Classifier} architecture \\ \hline
    Input $61 \times 64$ image \\
    GlobalMaxPooling \\
    Linear $(64,32)$ \& LayerNorm \& ReLU \\ 
    Linear $(32,3)$ \&  LogSoftmax \\ 
    Output $1 \times 3$ log. probabilities \\
    \hline
    \end{tabular}
\end{center}
\end{table}

EEGNet (see Table 2 in [19]) has effectively 3 hyperparameters: $F_1$ - number of temporal filters, $F_2$ - number of pointwise filters, $D$ - number of spatial filters. The architecture with $F_1 = 2$, $D = 2$, $F_2 = 4$ achieved the best classification accuracy on the test dataset.

ShallowNet (see Table 6 in [19]) has 5 hyperparameters: $F_1$ - number of temporal filters, $F_2$ - number of spatial filters, $D$ - filter size, $P$ - pooling kernel size and stride. The architecture with $F_1 = 20$, $F_2 = 30$, $D = 20$, $P = (20, 30)$ performed the best on the test dataset.